\documentstyle[12pt]{article}
 
\begin{document} 

\baselineskip 22pt 

\begin{center}
{\Large
\bf
Perturbative QCD Analysis\\
of $B$ to $\pi$ and $B$ to $\rho$ Transitions}\\
\vspace{1.0cm}
Dae Sung Hwang${}^{(a)}$ and Bum-Hoon Lee${}^{(b)}$\\
{\it{a: Department of Physics, Sejong University, Seoul 143--747,
Korea}}\\
{\it{b: Department of Physics, Sogang University, Seoul 121--742,
Korea}}\\
\vspace{2.0cm}
{\bf Abstract}\\
\end{center}

We calculate the form factors of $B\rightarrow\pi$
and $B\rightarrow\rho$ heavy to light transition matrix elements
by using the factorization formalism of perturbative QCD.
We obtain them at $q^2=0$ and show their dependences on the parameter
$\epsilon$ of the B meson distribution amplitude.
We also obtain the form factors as functions of $q^2$
in the region $0\le q^2 \le M_B^2/2$.
The relations among the form factors are found
in the limit of $m_\pi /M_B=0$, $m_\rho /M_B=0$, ${M_b/M_B}=1$ and $(1-x)<<1$.
\\

\vfill 


\vspace*{0.5cm}

\noindent
$^a$e-mail: dshwang@cs.sejong.ac.kr\\
$^b$e-mail: bhl@ccs.sogang.ac.kr\\

\thispagestyle{empty} 
\pagebreak 

\baselineskip 22pt

\noindent
{\bf \large 1. Introduction}\\

CP-violation is one of the most important and mysterious phenomena
in high energy physics,
for which we have only the
$K_L\rightarrow \pi\pi$ decay \cite{christen}
and the charge asymmetry in the decay
$K_L\rightarrow {\pi}^{\pm} l^{\mp} \nu$ \cite{dorfan}
for more than 30 years.
The mechanism of CP-violation through the complex phase of the
Cabibbo-Kobayashi-Maskawa (CKM) \cite{ckm} three family mixing matrix
in the Weinberg-Salam
model is presently the standard model for CP-violation.
The B meson system offers many possibilities to investigate CP-violation
\cite{sanda}, and the B-factories in KEK and SLAC are under construction
for this purpose.
In order to probe the CKM model precisely,
it is crucial obtain the values of the CKM matrix elements accurately
from B meson decays.
For the decays involving $b\rightarrow c$ transition,
we can apply the heavy quark
symmetry and it is possible to determine $V_{cb}$ reliably through the
heavy quark effective theory (HQET) \cite{iw}.
However, for those involving $b\rightarrow u$ it is less likely that
the heavy quark symmetry applies, and the determination of $V_{ub}$
has heavily relied on the models for the form factors.

The dynamical content of hadron decays is described
by Lorentz invariant form factors of current matrix elements.
The theoretical calculation of the form factors involving $b\rightarrow u$
transition is a difficult task,
since it is concerned with the nonperturbative realm of QCD
and we cannot apply the heavy quark symmetry.
Recently there have been active investigations of the form factors of
$B\rightarrow\pi$ and $B\rightarrow\rho$ by using quark model, QCD sum rule
and lattice calculations \cite{wittig}.
CLEO has presented first experimental results of the branching ratios of
$B\rightarrow\pi l\nu$ and $B\rightarrow\rho l\nu$ \cite{cleo1},
which are still model dependent.

In this paper we will calculate the form factors of $B\rightarrow\pi$
($F_0$, $F_1$) and $B\rightarrow\rho$ transitions ($V$, $A_0$, $A_1$, $A_2$)
by using the method which Szczepaniak et al. employed
for obtaining the $B\rightarrow\pi$ form factors \cite{shb}.
This method is based on the meson theory of Brodsky and Lepage \cite{bl}.
Ref. \cite{shb} noticed that in the case of a heavy meson decaying into two
lighter mesons the large momentum transfers are involved and the factorization
formula of perturbative QCD (PQCD) for exclusive reactions becomes applicable:
the amplitude can be written as a convolution of a hard-scattering quark-gluon
amplitude $T_{\rm h}$ and mason distribution amplitudes $\phi (x,Q^2)$ which
describe the fractional longitudinal momentum distribution amplitude of the
quark and antiquark in each meson.

In the present work we calculate the form factors at $q^2=0$, where $q^\mu$
is the difference between initial and final meson momenta, for various values of
$\epsilon$ which is a parameter describing the sharpness of the initial heavy
meson distribution amplitude.
Therefore we show the $\epsilon$ dependences of the form factors, which in return
can also be useful for understanding the structure of the B meson.
Then we obtain the $q^2$ dependences of the form factors in the region
$0\le q^2 \le M_B^2/2$, since it can be considered that in this region
the large momentum transfers are involved for the interaction between the quark
and antiquark in the meson.
We also obtain the relations among the form factors of $B\rightarrow\pi$ and
$B\rightarrow\rho$ transitions
in the limit of $m_\pi /M_B=0$, $m_\rho /M_B=0$, ${M_b/M_B}=1$ and $(1-x)<<1$.
We think these relations are valuable for improving the knowledge of the heavy
to light transition form factors.

In section 2 we study the form factors of $B\rightarrow\pi$, $F_0^{B\pi}(q^2)$
and $F_1^{B\pi}(q^2)$. In section 3, those of $B\rightarrow\rho$,
$V^{B\rho}(q^2)$, $A_1^{B\rho}(q^2)$, $A_2^{B\rho}(q^2)$ and $A^{B\rho}(q^2)$,
are calculated.
We obtain in section 4 the form factors and the relations among them
in the limit of $m_\pi /M_B=0$, $m_\rho /M_B=0$, ${M_b/M_B}=1$ and $(1-x)<<1$.
We organize our results of the form factors and compare them with
other existing calculations.
Section 5 constitutes the conclusion.\\

\noindent
{\bf \large 2. Form Factors $F_0^{B\pi}(q^2)$ and
$F_1^{B\pi}(q^2)$}\\

{}From Lorentz invariance one finds the decomposition of the
hadronic matrix element in terms of hadronic form factors
\cite{wsb}:
\begin{eqnarray}
& &<\pi^-(p_\pi )|V^\mu |B^0(p_B)>
\nonumber\\
&=&(p_B+p_\pi )^\mu f_+^{B\pi}(q^2)
+ (p_B-p_\pi )^\mu f_-^{B\pi}(q^2)
\nonumber\\
&=&\Bigl( r^\mu
-{m_B^2-m_{\pi}^2\over q^2}q^\mu \Bigr) \, F_1^{B\pi}(q^2)
+{m_B^2-m_{\pi}^2\over q^2}\, q^\mu \, F_0^{B\pi}(q^2),
\label{h1}
\end{eqnarray}
where $V^\mu = {\bar{u}}\gamma^\mu b$,
$q^\mu =(p_B-p_\pi )^\mu$, $r^\mu =(p_B+p_\pi )^\mu$, and
\begin{equation}
F_1^{B\pi}(q^2)=f_+^{B\pi}(q^2),\qquad
F_0^{B\pi}(q^2)=f_+^{B\pi}(q^2)
+{q^2\over M_B^2-m_{\pi}^2}f_-^{B\pi}(q^2).
\label{z1}
\end{equation}
In the rest frame of the decay products, $F_1$ and $F_0$
correspond to $1^-$ and $0^+$ exchanges, respectively.
At $q^2=0$ we have the constraint
\begin{equation}
F_1^{B\pi}(0)=F_0^{B\pi}(0),
\label{h2}
\end{equation}
since the hadronic matrix element in (\ref{h1}) is nonsingular
at this kinematic point.

We calculate the $B$ to $\pi$ (heavy to light) transition matrix
element by using the PQCD factorization of exclusive amplitudes
at high momentum transfer and neglect all final state interactions
\cite{shb}.
To the first order in $\alpha_{\rm s}=\alpha_{\rm s}(Q^2)$ we have
\begin{eqnarray}
& &<\pi^-(p_\pi )|V^\mu |B^0(p_B)>
={8\pi \alpha_{\rm s} \over 3}\,
\int_0^1dx\, \int_0^{1-\epsilon} dy\, \phi_B (x)\,
\nonumber\\
&\times &
[{{\rm Tr}\{ ({\not{p_\pi}}+m_\pi ) \gamma_5\gamma^\nu
{\not{k_1}}\gamma^\mu
({\not{p_B}}+g(x)M_B)\gamma_5\gamma_\nu\}\over k_1^2Q^2}\,
\nonumber\\
& & +
{{\rm Tr}\{ ({\not{p_\pi}}+m_\pi ) \gamma_5
\gamma^\mu ({\not{k_2}}+M_b)
\gamma^\nu
({\not{p_B}}+g(x)M_B)\gamma_5\gamma_\nu\}\over (k_2^2-M_b^2)Q^2}]\,
\phi_\pi (y) ,
\label{r3z}
\end{eqnarray}
where $Q^\mu =(1-x)p_B^\mu -(1-y)p_{\pi}^\mu$,
$k_1^\mu =-(1-x)p_B^\mu +p_{\pi}^\mu$,
$k_2^\mu =p_B^\mu -(1-y)p_{\pi}^\mu$, and
\begin{eqnarray}
Q^2&=&M_B^2[-(1-x)(1-y)(1-{q^2\over M_B^2})+(1-x)^2
+( (1-y)^2-(1-x)(1-y) ) {m_{\pi}^2\over M_B^2}],
\nonumber\\
k_1^2&=&M_B^2[-(1-x)(1-{q^2\over M_B^2})+(1-x)^2
+( 1-(1-x) ) {m_{\pi}^2\over M_B^2}],
\nonumber\\
k_2^2-M_b^2&=&M_B^2[-(1-y)(1-{q^2\over M_B^2})+(1-{M_b^2\over M_B^2})
-(1-y)y{m_{\pi}^2\over M_B^2}].
\label{r4}
\end{eqnarray}
In (\ref{r3z}) we use the distribution amplitude given by
\cite{shb,bl,bj,carlson}
\begin{eqnarray}
&&\phi_{\pi}(x)={\sqrt{{3\over 2}}}\, f_{\pi}x(1-x),
\label{h6p}\\
&&\phi_B(x)={1\over 2{\sqrt{6}}}\, f_B
{\varphi (x)\over \int_0^1 \varphi (x)dx},\qquad
\varphi (x)={x^2(1-x)^2\over [\epsilon^2x+(1-x)^2]^2},
\label{h6}
\end{eqnarray}
whose integrals are related to the meson decay constant by
\begin{equation}
\int_0^1dx\, \phi_M(x)={1\over 2{\sqrt{6}}}\, f_M.
\label{h7}
\end{equation}
In the right hand sides of (\ref{h6}) and (\ref{h7}) there are
extra factor ${1\over {\sqrt{2}}}$ compared with \cite{shb}, since
in this paper we adopt the convention of the meson decay constant
given by $<0|A^\mu |M(p)>=if_Mp^\mu$ in which
$f_{\pi}\equiv f_{\pi^+}=131.74\pm 0.15$ MeV \cite{rpp}.
In (\ref{r3z}) we took the upper limit of the integration over momentum
fraction $y$ of a quark in the light meson as ${1-\epsilon}$,
since the integration in
the interval ${1-\epsilon}\le y\le 1$ corresponds to the
Drell-Yan-West \cite{drell} end-point region. It gives only a small correction
to the form factors, and this region is also expected to be suppressed by
a Sudakov form factor \cite{shb}.

After some calculations we have
\begin{equation}
<\pi^-(p_\pi )|V^\mu |B^0(p_B)>=
{8\pi \alpha_{\rm s} \over 3}\,
\int_0^1dx\, \int_0^{1-\epsilon} dy\, \phi_B (x)\,
[{{\bar{K}}^a\over  k_1^2Q^2}
+{{\bar{K}}^b\over  (k_2^2-M_b^2)Q^2}]\,
\phi_\pi (y) ,
\label{r3zaa}
\end{equation}
where
\begin{eqnarray}
& &{\bar{K}}^a=4M_B^2
\{
r^\mu [-(1-x){q^2\over M_B^2}-x2g{m_\pi\over M_B}
-x{m_{\pi}^2\over M_B^2}]
\nonumber\\
& + &q^\mu [(1-x)(2-{q^2\over M_B^2})+(2-x)2g{m_\pi\over M_B}
-x{m_{\pi}^2\over M_B^2}]\} ,
\label{z2aa}\\
& &{\bar{K}}^b=4M_B^2
\{
r^\mu [(2g{M_b\over M_B}-1)+(1-y)(1-{q^2\over M_B^2})
+({M_b\over M_B}-y2g){m_\pi\over M_B}]
\nonumber\\
& + &q^\mu [-(2g{M_b\over M_B}-1)-(1-y)(1-{q^2\over M_B^2})
+({M_b\over M_B}-(2-y)2g){m_\pi\over M_B}
-2(1-y){m_{\pi}^2\over M_B^2}]\} .
\nonumber
\end{eqnarray}
Then from (\ref{h1}) and (\ref{r3zaa}) we have
\begin{eqnarray}
F_1^{B\pi}(q^2)
&=&
{8\pi \alpha_{\rm s} \over 3}\,
\int_0^1dx\, \int_0^{1-\epsilon} dy\, \phi_B (x)\,
[{{\bar{F}}_1^a\over  k_1^2Q^2}
+{{\bar{F}}_1^b\over  (k_2^2-M_b^2)Q^2}]\,
\phi_\pi (y) ,
\label{z4aa}\\
{\bar{F}}_1^a&=&4M_B^2[-(1-x){q^2\over M_B^2}-x2g{m_\pi\over M_B}
-x{m_{\pi}^2\over M_B^2}],
\nonumber\\
{\bar{F}}_1^b&=&4M_B^2[(2g{M_b\over M_B}-1)+(1-y)(1-{q^2\over M_B^2})
+({M_b\over M_B}-y2g){m_\pi\over M_B}],
\nonumber
\end{eqnarray}
and
\begin{eqnarray}
F_0^{B\pi}(q^2)
&=&
{8\pi \alpha_{\rm s} \over 3}\,
\int_0^1dx\, \int_0^{1-\epsilon} dy\, \phi_B (x)\,
[{{\bar{F}}_0^a\over  k_1^2Q^2}
+{{\bar{F}}_0^b\over  (k_2^2-M_b^2)Q^2}]\,
\phi_\pi (y),
\label{z5aa}\\
{\bar{F}}_0^a&=&4M_B^2[[-(1-x){q^2\over M_B^2}-x2g{m_\pi\over M_B}
-x{m_{\pi}^2\over M_B^2}]
\nonumber\\
&+&{q^2\over M_B^2-m_{\pi}^2}
[(1-x)(2-{q^2\over M_B^2})+(2-x)2g{m_\pi\over M_B}
-x{m_{\pi}^2\over M_B^2}]],
\nonumber\\
{\bar{F}}_0^b&=&4M_B^2[[(2g{M_b\over M_B}-1)+(1-y)(1-{q^2\over M_B^2})
+({M_b\over M_B}-y2g){m_\pi\over M_B}]
\nonumber\\
&+&{q^2\over M_B^2-m_{\pi}^2}
[-(2g{M_b\over M_B}-1)-(1-y)(1-{q^2\over M_B^2})
\nonumber\\
& &\qquad\qquad\qquad +({M_b\over M_B}-(2-y)2g){m_\pi\over M_B}
-2(1-y){m_{\pi}^2\over M_B^2}]].
\nonumber
\end{eqnarray}

For $m_\pi =0$, we have
\begin{eqnarray}
Q^2&=&M_B^2[-(1-x)(1-y)(1-{q^2\over M_B^2})+(1-x)^2],
\nonumber\\
k_1^2&=&M_B^2[-(1-x)(1-{q^2\over M_B^2})+(1-x)^2],
\nonumber\\
k_2^2-M_b^2&=&M_B^2[-(1-y)(1-{q^2\over M_B^2})+(1-{M_b^2\over M_B^2})],
\label{zr4a}
\end{eqnarray}
\begin{eqnarray}
{\bar{F}}_1^a&=&4M_B^2[-(1-x){q^2\over M_B^2}],
\label{z7aa}\\
{\bar{F}}_1^b&=&4M_B^2[(2g{M_b\over M_B}-1)+(1-y)(1-{q^2\over M_B^2})],
\nonumber
\end{eqnarray}
and
\begin{eqnarray}
{\bar{F}}_0^a&=&4M_B^2[(1-x){q^2\over M_B^2}(1-{q^2\over M_B^2})],
\label{z8aa}\\
{\bar{F}}_0^b&=&4M_B^2[[(2g{M_b\over M_B}-1)+(1-y)(1-{q^2\over M_B^2})]
(1-{q^2\over M_B^2})].
\nonumber
\end{eqnarray}

Then, for $m_\pi =0$ and ${M_b\over M_B}=1$, we have
\begin{equation}
F_{1,0}^{B\pi}(q^2)
={32\pi \alpha_{\rm s} \over 3M_B^2}\,
\int_0^1dx\, \int_0^{1-\epsilon} dy\, \phi_B (x)\,\,
\phi_{\pi} (y){1 \over (1-x) (1-y)^2} f_{1,0},
\label{zrn1}
\end{equation}
where
\begin{eqnarray}
f_1 &=& 2(1-y){1 \over 1-{q^2 \over M_B^2}}
+[(2g-1)-(1-y)]{1 \over (1-{q^2 \over M_B^2})^2},
\label{zrn2}\\
f_0 &=& [(2g-1)+(1-y)]{1 \over 1-{q^2 \over M_B^2}}.
\label{zrn3}
\end{eqnarray}
\\

\noindent
{\bf \large 3. Form Factors $V^{B\rho}(q^2)$, $A_1^{B\rho}(q^2)$,
$A_2^{B\rho}(q^2)$ and $A^{B\rho}(q^2)$}\\

{}From Lorentz invariance one finds the decomposition of the
hadronic matrix element in terms of hadronic form factors
\cite{wsb}:
\begin{eqnarray}
& &<\rho^-(p_\rho ,\varepsilon )|(V-A)^\mu |B^0(p_B)>
={2V(q^2)\over M_B+m_\rho}i\varepsilon_{\mu\alpha\beta\gamma}
\varepsilon^{*\alpha}p_B^\beta p_{\rho}^{\gamma}
\label{r1}\\
& &-(M_B+m_\rho )\varepsilon^{*\mu}A_1(q^2)
+{(\varepsilon^{*}\cdot p_B)\over M_B+m_\rho}(p_B+p_{\rho})^\mu
A_2(q^2)-2m_\rho{(\varepsilon^{*}\cdot p_B)\over q^2}q^\mu A(q^2).
\nonumber
\end{eqnarray}
The form factor $A(q^2)$ can be written as
\begin{equation}
A(q^2)=A_0(q^2)-A_3(q^2),\ \ \ {\rm where}\ \
A_3(q^2)={M_B+m_\rho \over 2m_\rho}A_1(q^2)
-{M_B-m_\rho \over 2m_\rho}A_2(q^2),
\label{r2}
\end{equation}
and at $q^2=0$ we have the constraint
\begin{equation}
A_0(0)=A_3(0).
\label{r2r}
\end{equation}

We calculate the $B$ to $\rho$ (heavy to light) transition matrix
element by using the PQCD factorization of exclusive amplitudes
at high momentum transfer and neglect all final state interactions
\cite{shb}.
To the first order in $\alpha_{\rm s}=\alpha_{\rm s}(Q^2)$ we have
\begin{eqnarray}
& &<\rho^-(p_\rho ,\varepsilon )|V^\mu |B^0(p_B)>
={8\pi \alpha_{\rm s} \over 3}\, 
\int_0^1dx\, \int_0^{1-\epsilon} dy\, \phi_B (x)\,
\nonumber\\
&\times &
[{{\rm Tr}\{ ({\not{p_\rho}}+m_\rho ) {\not{\varepsilon}}\gamma^\nu
{\not{k_1}}\gamma^\mu 
({\not{p_B}}+g(x)M_B)\gamma_5\gamma_\nu\}\over k_1^2Q^2}\,
\nonumber\\
& & + 
{{\rm Tr}\{ ({\not{p_\rho}}+m_\rho ) {\not{\varepsilon}}
\gamma^\mu ({\not{k_2}}+M_b)
\gamma^\nu
({\not{p_B}}+g(x)M_B)\gamma_5\gamma_\nu\}\over (k_2^2-M_b^2)Q^2}]\,
\phi_\rho (y) ,
\label{r3}
\end{eqnarray}
where $V^\mu = {\bar{u}}\gamma^\mu b$,
$Q^\mu =(1-x)p_B^\mu -(1-y)p_{\rho}^\mu$,
$k_1^\mu =-(1-x)p_B^\mu +p_{\rho}^\mu$,
$k_2^\mu =p_B^\mu -(1-y)p_{\rho}^\mu$, and
$Q^2$, $k_1^2$ and $k_2^2-M_b^2$ are given by (\ref{r4}) with $m_{\pi}$
replaced by $m_{\rho}$.
In (\ref{r3}) we use the distribution amplitude of B meson given in
(\ref{h6}) and that of $\rho$ meson given by
\cite{shb,bl,bj,carlson}
\begin{equation}
\phi_{\rho}(x)={\sqrt{{3\over 2}}}\, f_{\rho}x(1-x),
\label{h6a}
\end{equation}
where $<0|V^\mu |\rho (\varepsilon )>=f_{\rho}m_{\rho}{\varepsilon}^{\mu}$
in which
$f_{\rho}\equiv f_{\rho^+}=216\pm 5$ MeV \cite{rpp}.

After some calculations we have
\begin{equation}
<\rho^-(p_\rho ,\varepsilon )|V^\mu |B^0(p_B)>=
{8\pi \alpha_{\rm s} \over 3}\,
\int_0^1dx\, \int_0^{1-\epsilon} dy\, \phi_B (x)\,
[{{\bar{V}}^a\over  k_1^2Q^2}
+{{\bar{V}}^b\over  (k_2^2-M_b^2)Q^2}]\,
\phi_\rho (y) ,
\label{r4v1}
\end{equation}
where
\begin{eqnarray}
{\bar{V}}^a&=&8M_B{m_\rho\over M_B}
i\varepsilon_{\mu\alpha\beta\gamma}
\varepsilon^{*\alpha}p_B^\beta p_{\rho}^{\gamma},
\nonumber\\
{\bar{V}}^b&=&
8M_B(-(2g-{M_b\over M_B})-(1-y){m_\rho\over M_B})
i\varepsilon_{\mu\alpha\beta\gamma}
\varepsilon^{*\alpha}p_B^\beta p_{\rho}^{\gamma},
\label{r5}
\end{eqnarray}
\begin{eqnarray}
& &[{{\bar{V}}^a\over k_1^2Q^2}
+{{\bar{V}}^b\over (k_2^2-M_b^2)Q^2}]=
8M_Bi\varepsilon_{\mu\alpha\beta\gamma}
\varepsilon^{*\alpha}p_B^\beta p_{\rho}^{\gamma}
\nonumber\\
&\times &[{1\over k_1^2Q^2}{m_\rho\over M_B}+
{1\over (k_2^2-M_b^2)Q^2}(-(2g-{M_b\over M_B})-(1-y){m_\rho\over M_B})].
\label{r6}
\end{eqnarray}

To the first order in $\alpha_{\rm s}=\alpha_{\rm s}(Q^2)$ we have
\begin{eqnarray}
& &<\rho^-(p_\rho ,\varepsilon )|A^\mu |B^0(p_B)>
={8\pi \alpha_{\rm s} \over 3}\,
\int_0^1dx\, \int_0^{1-\epsilon} dy\, \phi_B (x)\,
\nonumber\\
&\times &
[{{\rm Tr}\{ ({\not{p_\rho}}+m_\rho ) {\not{\varepsilon}}\gamma^\nu
{\not{k_1}}\gamma^\mu\gamma_5
({\not{p_B}}+g(x)M_B)\gamma_5\gamma_\nu\}\over k_1^2Q^2}\,
\nonumber\\
& & +
{{\rm Tr}\{ ({\not{p_\rho}}+m_\rho ) {\not{\varepsilon}}
\gamma^\mu\gamma_5 ({\not{k_2}}+M_b)
\gamma^\nu
({\not{p_B}}+g(x)M_B)\gamma_5\gamma_\nu\}\over (k_2^2-M_b^2)Q^2}]\,
\phi_\rho (y) ,
\end{eqnarray}
where $A^\mu = {\bar{u}}\gamma^\mu\gamma_5 b$.
After some calculations we have
\begin{equation}
<\rho^-(p_\rho ,\varepsilon )|A^\mu |B^0(p_B)>=
{8\pi \alpha_{\rm s} \over 3}\,
\int_0^1dx\, \int_0^{1-\epsilon} dy\, \phi_B (x)\,
[{{\bar{A}}^a\over  k_1^2Q^2}
+{{\bar{A}}^b\over  (k_2^2-M_b^2)Q^2}]\,
\phi_\rho (y) ,
\label{r4v2}
\end{equation}
where
\begin{eqnarray}
{\bar{A}}^a&=&\varepsilon^{*\mu} M_B^24m_\rho 
(-(1-{q^2\over M_B^2})+2(1-x)-{m_{\rho}^2\over M_B^2})
\nonumber\\
&+&(\varepsilon^{*}\cdot p_B)r^\mu 4m_\rho (1-2(1-x))
\nonumber\\
&+&(\varepsilon^{*}\cdot p_B)q^\mu 4m_\rho (-1-2(1-x)),
\label{r7}\\
{\bar{A}}^b&=&\varepsilon^{*\mu} 4M_B^3
[((2g-{M_b\over M_B})-(1-y){m_\rho\over M_B})(1-{q^2\over M_B^2})
\nonumber\\
& &\ \ \ -2(2g{M_b\over M_B}-1){m_\rho\over M_B}
+((2g-{M_b\over M_B})-4g(1-y)){m_{\rho}^2\over M_B^2}
-(1-y){m_{\rho}^3\over M_B^3}]
\nonumber\\
&+&(\varepsilon^{*}\cdot p_B)r^\mu 
4M_B(-(2g-{M_b\over M_B})-(1-y){m_\rho\over M_B})
\nonumber\\
&+&(\varepsilon^{*}\cdot p_B)q^\mu
4M_B(-1)(-(2g-{M_b\over M_B})-(1-y){m_\rho\over M_B}),
\nonumber
\end{eqnarray}

\begin{eqnarray}
& &[{{\bar{A}}^a\over k_1^2Q^2}
+{{\bar{A}}^b\over (k_2^2-M_b^2)Q^2}]=
4M_B^3
\nonumber\\
&\times &\{ \varepsilon^{*\mu}[
{1\over k_1^2Q^2}{m_\rho\over M_B}
(-(1-{q^2\over M_B^2})+2(1-x)-{m_{\rho}^2\over M_B^2})
\nonumber\\
& &\ \ \ \ \ +{1\over (k_2^2-M_b^2)Q^2}
[((2g-{M_b\over M_B})-(1-y){m_\rho\over M_B})(1-{q^2\over M_B^2})
\nonumber\\
& &\ \ \ \ \ -2(2g{M_b\over M_B}-1){m_\rho\over M_B}
+((2g-{M_b\over M_B})-4g(1-y)){m_{\rho}^2\over M_B^2}
-(1-y){m_{\rho}^3\over M_B^3}]]
\nonumber\\
& &+{(\varepsilon^{*}\cdot p_B)r^\mu\over M_B^2}[
{1\over k_1^2Q^2}{m_\rho\over M_B}
(1-2(1-x))
\nonumber\\
& &\ \ \ \ \ +{1\over (k_2^2-M_b^2)Q^2}
(-(2g-{M_b\over M_B})-(1-y){m_\rho\over M_B})]
\nonumber\\
& &+{(\varepsilon^{*}\cdot p_B)q^\mu\over M_B^2}[
{1\over k_1^2Q^2}{m_\rho\over M_B}
(-1-2(1-x))
\nonumber\\
& &\ \ \ \ \ +{1\over (k_2^2-M_b^2)Q^2}
(-1)(-(2g-{M_b\over M_B})-(1-y){m_\rho\over M_B})]\}.
\label{r8}
\end{eqnarray}

For $m_\rho =0$,
$Q^2$, $k_1^2$ and $k_2^2-M_b^2$ are given by (\ref{zr4a}), and
we have
\begin{eqnarray}
& &[{{\bar{V}}^a\over k_1^2Q^2}
+{{\bar{V}}^b\over (k_2^2-M_b^2)Q^2}]=
{{\bar{V}}^b\over (k_2^2-M_b^2)Q^2}
\nonumber\\
&=&
8M_Bi\varepsilon_{\mu\alpha\beta\gamma}
\varepsilon^{*\alpha}p_B^\beta p_{\rho}^{\gamma}
{1\over (k_2^2-M_b^2)Q^2}(2g-{M_b\over M_B})(-1),
\label{r9}
\end{eqnarray}
and
\begin{eqnarray}
& &[{{\bar{A}}^a\over k_1^2Q^2}
+{{\bar{A}}^b\over (k_2^2-M_b^2)Q^2}]
={{\bar{A}}^b\over (k_2^2-M_b^2)Q^2}
\nonumber\\
&=&
4M_B^3
{1\over (k_2^2-M_b^2)Q^2}(2g-{M_b\over M_B})
\nonumber\\
&\times &
\{ \varepsilon^{*\mu}(1-{q^2\over M_B^2})
+{(\varepsilon^{*}\cdot p_B)r^\mu\over M_B^2}(-1)
+{(\varepsilon^{*}\cdot p_B)q^\mu\over M_B^2}\} .
\label{r10}
\end{eqnarray}
\\

\noindent
{\bf \large 4. Relations among Form Factors in the Limit of
$m_\pi /M_B=0$,\\
\hspace*{0.5cm} $m_\rho /M_B=0$, ${M_b/M_B}=1$ and $(1-x)<<1$}\\

In this section we study the form factors $F_0^{B\pi}(q^2)$ and
$F_1^{B\pi}(q^2)$ of $B\rightarrow\pi$,
and $V^{B\rho}(q^2)$, $A_0^{B\rho}(q^2)$, $A_1^{B\rho}(q^2)$ and
$A_2^{B\rho}(q^2)$ of $B\rightarrow\rho$,
in the limit of $m_\pi /M_B=0$, $m_\rho /M_B=0$, ${M_b/M_B}=1$ and $(1-x)<<1$.
The approximations
$m_\pi /M_B=0$, $m_\rho /M_B=0$ and ${M_b/M_B}=1$ are reasonable ones,
since the B meson mass is much larger than the masses of light masons or
light quarks.
$(1-x)<<1$ is also a good approximation in the region $0\le q^2 \le M_B^2/2$
as can be seen from (\ref{r4}),
since $(1-x)$ is roughly given by the ratio of light and $b$ quark masses
or roughly by the value of the parameter $\epsilon$ in the B meson
distribution amplitude (\ref{h6}).

{}From (\ref{zrn1})$-$(\ref{r1}) and (\ref{r9})$-$(\ref{r10}), we can organize
the form factors
as follows:
\begin{equation}
F_i(q^2)
={32\pi \alpha_{\rm s} \over 3M_B^2}\,
\int_0^1dx\, \int_0^{1-\epsilon} dy\, \phi_B (x)\,\,
\phi_{\pi} (y){1 \over (1-x) (1-y)^2} f_i,
\label{rn1}
\end{equation}
where $F_i$$=$$F_0$,$F_1$,$V$,$A_0$,$A_1$,$A_2$,
and $\phi_{i} (y)=\phi_{\pi} (y)$ for $F_0$ and $F_1$,
and $\phi_{i} (y)=\phi_{\rho} (y)$ for $V$, $A_0$, $A_1$ and $A_2$.
In (\ref{rn1}) $f_i$ are given by
\begin{eqnarray}
f_0 &=& [(2g-1)+(1-y)]{1 \over z},
\label{rn2}\\
f_1 &=& 2(1-y){1 \over z}
+[(2g-1)-(1-y)]{1 \over z^2},
\label{rn3}\\
v &=& (-1)(2g-1){1 \over z^2},
\label{rn4}\\
a_1 &=& (2g-1){1 \over z},
\label{rn5}\\
a_2 &=& (2g-1){1 \over z^2},
\label{rn6}\\
a &=& {q^2 \over 2m_\rho M_B} (2g-1){1 \over z^2},
\label{rn7}
\end{eqnarray}
where $z\equiv 1-{q^2 \over M_B^2}$.
By taking the terms up to the first order in $m_\rho /M_B$
for $a_1$, $a_2$ and $a$ in (\ref{r7}) and (\ref{r8}),
we obtain from the relations (\ref{r2}):
\begin{equation}
a_0 = (-1)[(2g-1)+(1-y)]
{1 \over z^2}.
\label{rn8}
\end{equation}

{}From (\ref{rn1})$-$(\ref{rn8}) we find the relations among the form factors:
\begin{eqnarray}
&&F_1(q^2)=F_0(q^2)(2-{1\over z})+2{f_{\pi}\over f_{\rho}}
A_1(q^2)(-1+{1\over z})
\label{rn10}\\
&&F_0(q^2){1\over z}=-{f_{\pi}\over f_{\rho}}A_0(q^2)
\label{rn11}\\
&&A_1(q^2){1\over z}=A_2(q^2)=-V(q^2).
\label{rn12}
\end{eqnarray}
At $q^2=0$, we have the following relations:
\begin{equation}
F_1(0)=F_0(0)=-{f_{\pi}\over f_{\rho}}A_0(0),\qquad
A_1(0)=A_2(0)=-V(0).
\label{rn9}
\end{equation}
Ball and Braun also obtained the second relation in (\ref{rn9}) to their
accuracy in their QCD sum rule calculation \cite{bb}.
We calculate $F_1(0)$ and $A_1(0)$ from (\ref{rn1})$-$(\ref{rn2})
and (\ref{rn5}) for the values of the parameter $\epsilon$ of the B meson
distribution amplitude (\ref{h6}) in the range
$0.01\le\epsilon\le 0.10$, and present the results in Table 1 and Fig. 1.
In this calculation we took $g=1$, ${\alpha}_{\rm s}=0.38$ \cite{shb},
and $f_B=0.2$ GeV.
We find that $F_1(0)$ and $A_1(0)$ depend much on the value of $\epsilon$.
The commonly used value $F_1(0)=0.33$ obtained by Wirbel et al. \cite{wsb} in
quark model corresponds to $\epsilon =0.022$.
In this way, the information of the form factor is related to the structure
of the B meson, and those two help each other as clues for the
understanding of mesons.
For $\epsilon =0.022$ we have
\begin{equation}
F_1(0)=0.33,\qquad A_1(0)=0.47\qquad
{\rm for}\ \ \epsilon =0.022.
\label{rn13}
\end{equation}
The values of other form factors at $q^2=0$ can be given by the relations
in (\ref{rn9}).
In Table 1 and Fig. 1
we also present the dependence of the ratio $F_1(0)/A_1(0)$ on the parameter
$\epsilon$ in the range $0.01\le\epsilon\le 0.10$, and find that this ratio
is much less dependent on $\epsilon$.
In Table 2, we compare our results of the form factor values at $q^2=0$
given by (\ref{rn9}) and (\ref{rn13}) with other existing results
obtained by quark model, QCD sum rule and lattice calculations.

The $q^2$ dependences of the form factors are given by (\ref{rn1})$-$(\ref{rn8})
and (\ref{rn10})$-$(\ref{rn12}).
We obtain them in the region $0\le q^2 \le M_B^2/2$, and present the results
in Fig. 2.
The formulas in (\ref{rn1})$-$(\ref{rn8}) can be written as
\begin{eqnarray}
&&F_0(q^2)=(a+b){1\over z},\quad F_1(q^2)=2b{1\over z}+(a-b){1\over z^2},
\label{rn14}\\
&&A_1(q^2)=a{f_\rho\over f_\pi}{1\over z},\quad
A_2(q^2)=-V(q^2)=a{f_\rho\over f_\pi}{1\over z^2},\quad 
-A_0(q^2)=(a+b){f_\rho\over f_\pi}{1\over z^2},
\nonumber
\end{eqnarray}
where
\begin{eqnarray}
a&=&{32\pi \alpha_{\rm s} \over 3M_B^2}\,
\int_0^1dx\, \int_0^{1-\epsilon} dy\, \phi_B (x)\,\,
\phi_{\pi} (y){2g-1 \over (1-x) (1-y)^2},
\label{rn15}\\
b&=&{32\pi \alpha_{\rm s} \over 3M_B^2}\,
\int_0^1dx\, \int_0^{1-\epsilon} dy\, \phi_B (x)\,\,
\phi_{\pi} (y){1 \over (1-x) (1-y)}.
\nonumber
\end{eqnarray}
We note that the expressions in (\ref{rn14}), and also the relations
in (\ref{rn10})$-$(\ref{rn12}), are independent of the shapes of the distribution
amplitudes $\phi_B (x)$, $\phi_{\pi} (y)$ and the value of the parameter
$\epsilon$.
Their dependences appear in the values of $a$ and $b$ in (\ref{rn15}).
For the distribution amplitudes in (\ref{h6p}), (\ref{h6}) and (\ref{h6a}),
and $\epsilon =0.022$ which gives (\ref{rn13}),
we have $a=0.28$ and $b=0.05$.
{}From (\ref{rn14}) we find that $F_0(q^2)$ and $A_1(q^2)$ have the simple pole
$q^2$ dependence, and $A_2(q^2)$, $V(q^2)$ and $A_0(q^2)$ have the dipole
$q^2$ dependence.
$F_1(q^2)$ has the mixture of the simple pole and dipole $q^2$ dependences,
but the dipole $q^2$ dependence is dominant.
These characters of the form factors are shown clearly in Fig. 2.
\\

\pagebreak

\noindent
{\bf \large 5. Conclusion}\\

We calculated the form factors of $B\rightarrow\pi$
and $B\rightarrow\rho$ heavy to light transition matrix elements
by using the factorization formalism of perturbative QCD.
We obtained them at $q^2=0$
for the values of the parameter $\epsilon$ of the B meson
distribution amplitude in the range
$0.01\le\epsilon\le 0.10$,
and found that they depend much on the value of $\epsilon$
unless $g=1/2$.
We also obtained the $q^2$ dependences of the form factors in the region
$0\le q^2 \le M_B^2/2$, since we can consider that in this region
the large momentum transfers are involved for the interaction between the quark
and antiquark in the meson.
The relations among the form factors are found
in the limit of $m_\pi /M_B=0$, $m_\rho /M_B=0$, ${M_b/M_B}=1$ and $(1-x)<<1$.
These conditions are reasonable ones since
the B meson mass is much larger than the masses of light masons or
light quarks, and
$(1-x)$ is roughly given by the ratio of light and $b$ quark masses.

For the heavy to heavy transitions like $B\rightarrow D^{(*)}$, HQET can
be applied and all the relevant form factors are expressed by the one
Isgur-Wise function \cite{iw}.
However, for heavy to light transitions like $B\rightarrow \pi$ and
$B\rightarrow \rho$, we cannot apply HQET, and it is very important to
understand the form factors of heavy to light transitions better.
Improvements in this area of study are not only invaluable for the analyses of
experimental data, for example, in the extraction of the CKM matrix elements
from the experimental results of the B meson decay branching ratios, but also
for the better understanding of the structures of mesons.
Stech studied the form factors of heavy to light transitions in the latter
context \cite{stech}.
In the factorization formalism of perturbative QCD
we obtained the relations among the $q^2$ dependent form factors in
(\ref{rn10})$-$(\ref{rn12}), and the relations (\ref{rn9}) at $q^2=0$.
The second relaton $A_1(0)=A_2(0)=-V(0)$ in (\ref{rn9}) was also obtained by
Ball and Braun in their QCD sum rule calculation \cite{bb}.
In the first relation $F_1(0)=F_0(0)=-{f_{\pi}\over f_{\rho}}A_0(0)$
in (\ref{rn9}), the first
equality is a well-known relation as explained in (\ref{h2}), however,
the second equality is not a usual one.
This relation $F_1(0)=-{f_{\pi}\over f_{\rho}}A_0(0)$
can be checked by measuring the differential
branching ratios
$d{\cal B} (B^0\rightarrow \pi^- l^+ \nu)/dq^2$ and
$d{\cal B} (B^0\rightarrow \rho^- l^+ \nu)/dq^2$ at $q^2=0$, which are given by
\begin{eqnarray}
{d{\cal B} (B^0\rightarrow \pi^- l^+ \nu)\over dq^2}|_{q^2=0}&=&
{G_F^2M_B^5|V_{ub}|^2\over 192\pi^3\Gamma_B}(1-{m_{\pi}^2\over M_B^2})^3
|F_1(0)|^2,
\label{con1}\\
{d{\cal B} (B^0\rightarrow \rho^- l^+ \nu)\over dq^2}|_{q^2=0}&=&
{G_F^2M_B^5|V_{ub}|^2\over 192\pi^3\Gamma_B}(1-{m_{\rho}^2\over M_B^2})^3
|A_0(0)|^2.
\label{con2}
\end{eqnarray}
{}From (\ref{con1}) and (\ref{con2}) we have
\begin{equation}
{{d{\cal B} (B^0\rightarrow \pi^- l^+ \nu) / dq^2}|_{q^2=0}\over
{d{\cal B} (B^0\rightarrow \rho^- l^+ \nu) / dq^2}|_{q^2=0}}
={(1-{m_{\pi}^2\over M_B^2})^3\over (1-{m_{\rho}^2\over M_B^2})^3}
{|F_1(0)|^2\over |A_0(0)|^2}
=1.06\,\, {|F_1(0)|^2\over |A_0(0)|^2}.
\label{con3}
\end{equation}
CLEO reported \cite{cleo1}
${\cal B} (B^0\rightarrow \pi^- l^+ \nu)
=(1.8\pm 0.4\pm 0.3\pm 0.2)\times 10^{-4}$
and
${\cal B} (B^0\rightarrow \rho^- l^+ \nu)
=(2.5\pm 0.4\, {}^{+0.5}_{-0.7}\pm 0.5)\times 10^{-4}$.
Then we expect that the ratio in the left hand side of (\ref{con3}) will be
measured in near future, which will provide the ratio
$|F_1(0)|/|A_0(0)|$.

We obtained the expressions for the form factors given by (\ref{rn14})
and the relations among the form factors given by
(\ref{rn10})$-$(\ref{rn12}).
We note that they are independent of the shapes of the distribution
amplitudes $\phi_B (x)$, $\phi_{\pi} (y)$ and the value of the parameter
$\epsilon$.
Their dependences appear only in the numerical values of $a$ and $b$
in (\ref{rn15}).
The formulas in (\ref{rn14}) show that $F_0(q^2)$ and $A_1(q^2)$ have the
simple pole
$q^2$ dependence, and $A_2(q^2)$, $V(q^2)$ and $A_0(q^2)$ have the dipole
$q^2$ dependence.
$F_1(q^2)$ has the mixture of the simple pole and dipole $q^2$ dependences,
but the dipole $q^2$ dependence is dominant.
These results have been possible since in the case of the B meson decaying into
$\pi$ or $\rho$ meson with $q^2$ in the range of $0\le q^2 \le M_B^2/2$,
large momentum transfers are involved, and the factorization
formula of perturbative QCD for exclusive reactions becomes applicable.
Therefore, the heavy to light decays possess their own
characteristic and interesting properties
whose deeper understandings are desirable.
\\

\pagebreak

\vspace*{0.5cm}

\noindent
{\em Acknowledgements} \\
The authors are grateful to Stanley J. Brodsky for helpful discussions
and for reading the manuscript carefully.
They are also thankful to Adam Szczepaniak for useful discussions.
This work was supported
in part by the Basic Science Research Institute Program,
Ministry of Education, Project No. BSRI-97-2414,
in part by Korea Science and Engineering Foundation through the SRC
Program of SNU-CTP,
and in part by Non-Directed-Research-Fund,
Korea Research Foundation 1997.\\

\pagebreak

\pagebreak

\begin{table}[h]
\vspace*{0.5cm}
\hspace*{-0.8cm}
\begin{tabular}{|c|c|c|c|c|c|c|c|c|c|c|}   \hline
$\epsilon$&0.01&0.02&0.03&0.04&0.05&0.06&0.07&0.08&0.09&0.10\\
\hline
$F_1^{B\pi}(0)$&0.853&0.374&0.229&0.161&0.123&0.098&0.081&0.068&0.058&0.051\\
$A_1^{B\rho}(0)$&1.232&0.526&0.317&0.220&0.165&0.130&0.106&0.088&0.075&0.065\\
$F_1^{B\pi}(0)/A_1^{B\rho}(0)$&0.693&0.710&0.723&0.734&0.744
&0.754&0.762&0.771&0.779&0.786\\
\hline
\end{tabular}
\caption{The $\epsilon$ dependences of $F_1^{B\pi}(0)$, $A_1^{B\rho}(0)$ and
$F_1^{B\pi}(0)/A_1^{B\rho}(0)$.}
\end{table}

\begin{table}[h]
\vspace*{0.5cm}
\hspace*{-1.0cm}
\begin{tabular}{|c|c|c|c|c|}   \hline
&$F_1^{B\pi}(0)$&$A_1^{B\rho}(0)$
&$A_2^{B\rho}(0)$&$-V^{B\rho}(0)$\\
\hline
This work&0.33&0.47&0.47&0.47\\
\hline
(Quark Model)&&&&\\
WSB \cite{wsb}&0.33&0.28&0.28&0.33\\
ISGW \cite{isgw}&0.09&0.05&0.02&0.27\\
Jaus \cite{jaus}&0.27&0.26&0.24&0.35\\
FGM \cite{fgm}&$0.20\pm 0.02$&$0.26\pm 0.03$&$0.31\pm 0.03$&$0.29\pm 0.03$\\
Melikhov \cite{melikhov}&0.29&0.17$-$0.18&0.155&0.215\\
\hline
(QCD Sum Rule)&&&&\\
BKR \cite{bkr}&0.30&---&---&---\\
KRWY \cite{krwy}&0.27&---&---&---\\
BB \cite{bb}&---&$0.27\pm 0.05$&$0.28\pm 0.05$&$0.35\pm 0.07$\\
\hline
(Lattice)&&&&\\
UKQCD \cite{ukqcd}&$0.27\pm 0.11$&$0.27^{+5}_{-4}$&$0.25^{+5}_{-3}$
&$0.35^{+6}_{-5}$\\
GSS \cite{gss}&$0.43\pm 0.19$&$0.28\pm 0.03$&$0.46\pm 0.23$&$0.65\pm 0.15$\\
APE \cite{ape}&$0.35\pm 0.08$&$0.24\pm 0.12$&$0.27\pm 0.80$&$0.53\pm 0.31$\\
ELC \cite{elc}&$0.30\pm 0.14\pm 0.05$&$\ \ \ 0.22\pm 0.05\ \ \ $
&$0.49\pm 0.21\pm 0.05$&$\ \ \ 0.37\pm 0.11\ \ \ $\\
\hline
\end{tabular}
\caption{The results of this work of the form factor values at $q^2=0$
obtained with $\epsilon =0.022$, and
other existing results
obtained by quark model, QCD sum rule and lattice calculations.}
\end{table}

\pagebreak

\noindent
{\large\bf
Figure Captions}\\

\noindent
Fig. 1. The $\epsilon$ dependences of $F_1(0)$, $A_1(0)$ and
$F_1(0)/A_1(0)$.
\\

\noindent
Fig. 2. The $q^2$ dependences of the form factors.
$F_0(q^2)$ and $A_1(q^2)$ have the simple pole
dependence, and $A_2(q^2)$, $V(q^2)$ and $A_0(q^2)$ have the dipole
dependence.
$F_1(q^2)$ has the mixture of the simple pole and dipole dependences,
but the dipole dependence is dominant.

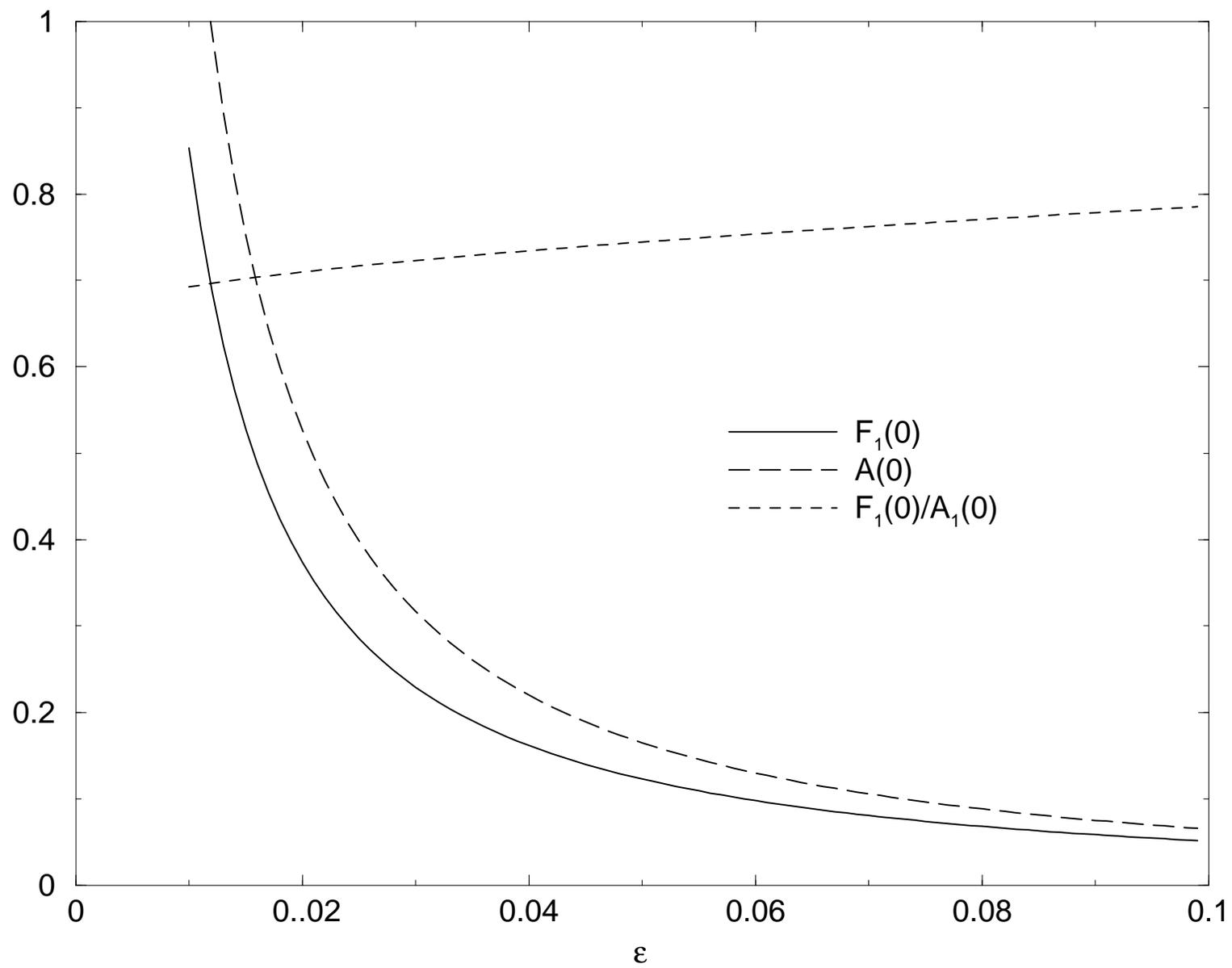
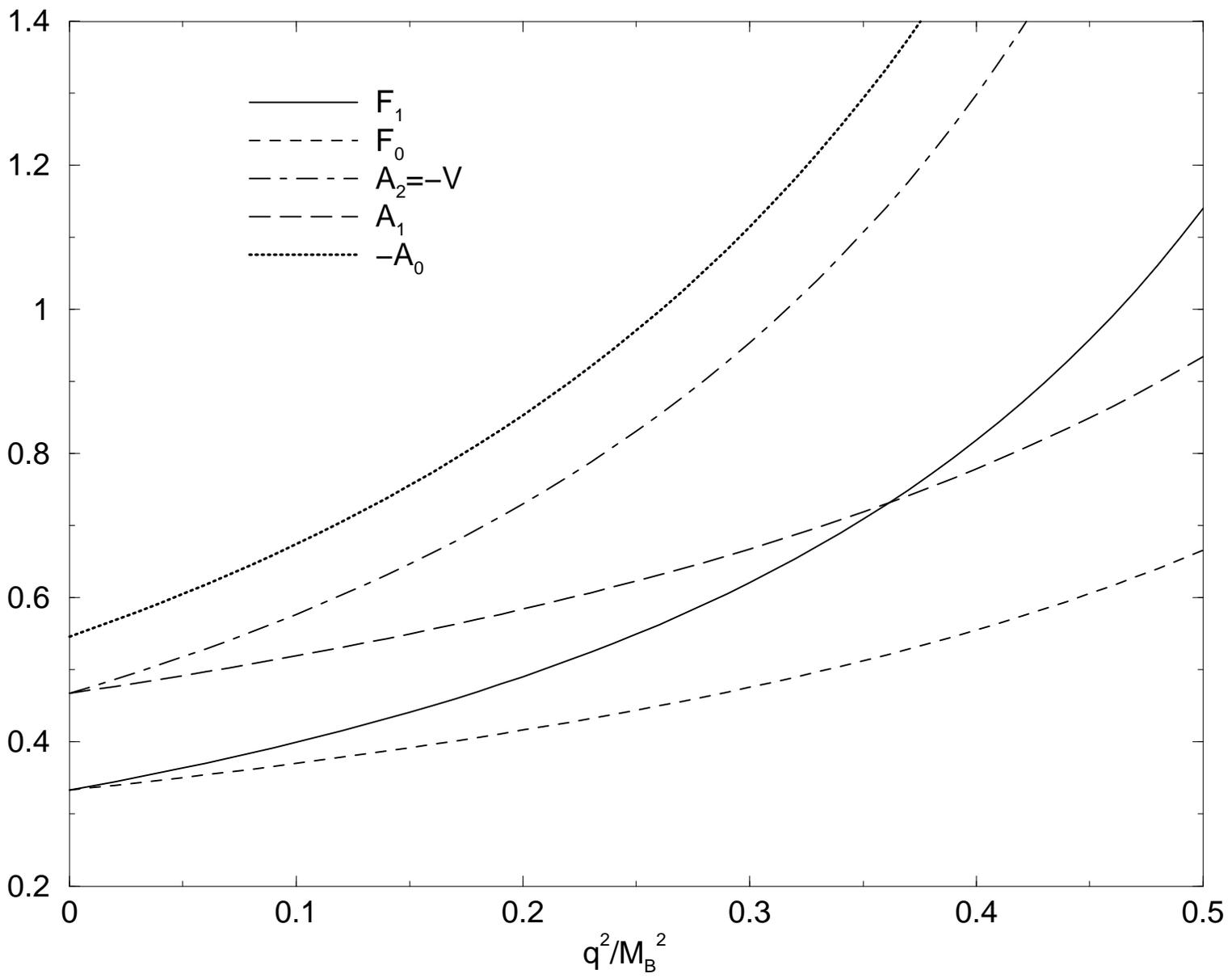
\end{document}